\documentclass[prd,preprint,nofootinbib]{revtex4-1}

\usepackage{amssymb,amsmath}
\usepackage{color}
\usepackage[toc,page]{appendix}
\usepackage{graphicx}
\usepackage{slashed}
\usepackage{tikz,subfig}
\usepackage{hyperref}
\usepackage{float} 
\usepackage[T1]{fontenc}

\renewcommand{\Im}{{\cal I}m}
\renewcommand{\Re}{{\cal R}e}

\newcommand{\D}{\mathrm{d}}

\newcommand{\ket}[1]{|{#1} \rangle}
\newcommand{\bra}[1]{\langle {#1}|}
\newcommand{\mean}[1]{\left\langle {#1}\right\rangle}
\newcommand{\bk}[2]{\langle{#1}|{#2}\rangle}
\newcommand{\mel}[3]{\langle{#2}|{#1}|{#3}\rangle}

\newcommand{\gmu}{\gamma^{\mu}}
\newcommand{\gnu}{\gamma^{\nu}}
\newcommand{\gsig}{\gamma^{\sigma}}
\newcommand{\grho}{\gamma^{\rho}}
\newcommand{\gMu}{\gamma_{\mu}}

\newcommand{\gfive}{\gamma^5}

\newcommand{\uspinor}[2]{u_{#1}(\vec{#2})}

\def\beq{\begin{equation}}
\def\eeq{\end{equation}}

\begin{document}
\title{Probing the two-neutrino exchange force using atomic parity violation}
\author{Mitrajyoti Ghosh}
\email{mg2338@cornell.edu}
\affiliation{Department of Physics, LEPP, Cornell University, Ithaca, NY 14853, USA}
\author{Yuval Grossman}
\email{yg73@cornell.edu}
\affiliation{Department of Physics, LEPP, Cornell University, Ithaca, NY 14853, USA}
\author{Walter Tangarife}
\email{wtangarife@luc.edu}
\affiliation{Department of Physics, Loyola University Chicago, Chicago, IL 60660, USA}

\begin{abstract}
The exchange of two neutrinos at one loop leads to a long-range parity-violating force between fermions. We explore the two-neutrino force in the backdrop of atomic physics. We point out that this is the largest parity-violating long-range force in the Standard Model and calculate the effect of this force in  experiments that probe atomic parity violation by measuring optical rotation of light as it passes through a sample of vaporized atoms. We perform explicit calculations for the hydrogen atom to demonstrate this effect. Although we find that the effect is too small to be observed in hydrogen in the foreseeable future, our approach may be applied to other setups where long-range parity violation is large enough to be probed experimentally. 
\end{abstract}

\maketitle

\tableofcontents

\section{Introduction}\label{intro}
\begin{figure}[t!]
\centerline{\includegraphics[width=4in]{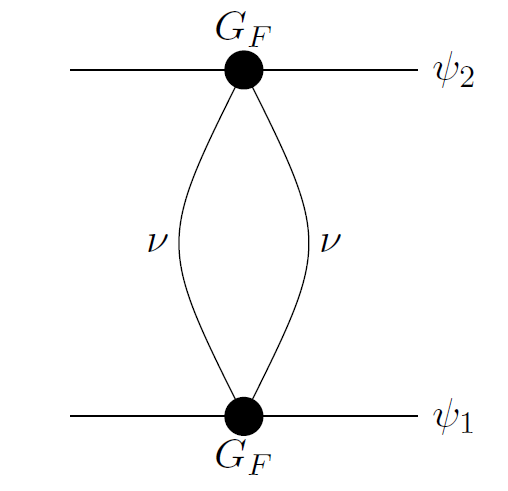}}
  \caption{The four-Fermi effective diagram for two-neutrino exchange forces between two fermions, labeled $\psi_1$ and $\psi_2$.}
  \label{fig:2nu}
\end{figure}

The fact that a pair of massless neutrinos mediate a long-range force via one-loop diagrams, as shown
in Fig.~\ref{fig:2nu}, has been
known for a long time \cite{Feinberg:1968zz,Feinberg:1989ps,Hsu:1992tg,Fischbach:1996qf}. At leading order, this diagram gives rise to a force of the form
\beq \label{eq:sim-nu-force}
V(r) = {G_F^2 \over 4\pi^3 r^5},
\eeq
where $G_F$ is the Fermi constant. The force is very weak. At distances larger than about a nanometer its magnitude is smaller that the gravitational force between two protons. At this scale, the electromagnetic Van der Waals force overpowers both. Thus, it has not been observed yet and furthermore, there is no realistic proposal to build an experiment that could see it. It is, therefore, an interesting question to ask if there is any way to probe this force that has not been explored yet. 

In many cases in the past, to observe a very small effect, one looked for symmetries that are broken by it. For example, the weak interaction was observed, even though it is much weaker than the strong and electromagnetic interactions, because it violates the flavor symmetries of these stronger forces. Thus, one way to try to achieve sensitivity to the two-neutrino force is to look for symmetries that it violates. 

In this paper, we point out that the two-neutrino force is the largest long-range parity-violating interaction in the Standard Model (SM). This is in contrast to the parity violation mediated by the $W$ and the $Z$ bosons, which is a short-distance effect. The reason is that in the case of the two-neutrino force the mediator is massless (or close to massless), while in the case of the $W$ and the $Z$ the mediators are massive.

In recent years atomic and molecular systems have attracted considerable interest as probes of physics within and beyond the SM. For instance, the work of Fichet \cite{Fichet:2017bng} explores molecular spectroscopy as a probe of dark matter. Another example is Ref.~\cite{Stadnik:2017yge}, where Stadnik shows how the long-range neutrino force can be probed using atomic and nuclear spectroscopy. Given that parity violation in atoms has also been suggested as a probe of new physics, for example, in \cite{Arcadi:2019uif}, a natural question to ask is whether it is possible to see effects of the neutrino force in parity-violation experiments done on atomic systems. In this paper, we explore this idea in some depth.

We find that the effect of the parity non-conserving force on atomic systems is tiny, much smaller than what one can hope to achieve in the near future. Yet, our approach in this paper can be used in other setups and, while we do not have a concrete idea where it can be practical, the hope is that a system where long-range parity violation can be large enough to probe experimentally will be found.

The arrangement of the paper is as follows: In Sec.~\ref{sec:neutrino_review}, we briefly review the literature regarding the two-neutrino force. Sec.~\ref{sec:APV_review} aims to provide some background on atomic parity violation. We discuss parity violating forces in atomic systems in Sec.~\ref{sec:pncforce}. Thereafter, we shift our focus to the hydrogen atom and compute the parity-violating two-neutrino force between the proton and the electron in the hydrogen atom in Sec.~\ref{neutrinoH}. The effects of this force on hydrogen eigenstates are discussed in Sec.~\ref{sec:effects}, while a sample calculation to illustrate the idea has been performed in Sec.~\ref{sec:sample_calc}. Finally, we present our concluding remarks in Sec.~\ref{sec:conclusion}. More details about the calculations in Sec.~\ref{neutrinoH} and Sec.~\ref{sec:sample_calc} are given in the Appendix.

\section{A review of the two-neutrino force}\label{sec:neutrino_review}

A classical force is mediated by a boson. The two-neutrino exchange gives rise to a long-range force since two fermions, to some extent, can be treated as a boson. This force is also called ``a quantum force'' as it arises at the loop level. In this section, we provide a brief review of the literature on the long-range force generated by the exchange of a pair of neutrinos.

Although the idea of a two-neutrino mediated force was conceived by Feynman \cite{Feynman:1996kb}, the first calculation of the force dates back to Ref.~\cite{Feinberg:1968zz}, where Feinberg and Sucher computed the leading form of the two-neutrino force to obtain Eq.~\eqref{eq:sim-nu-force}. They worked in the four-Fermi approximation, that is, neglecting terms of order $E/m_W$, $E$ being the energy of the interaction, and $m_W$ the mass of the $W$ boson. The same authors repeated the calculation in Ref.~\cite{Feinberg:1989ps} to incorporate the previously ignored neutral current interaction. In both calculations, the velocity-dependent terms of the potential were ignored under the assumption that the velocity of the fermions was much smaller than the speed of light. Later, Sikivie and Hsu performed a similar calculation in Ref.~\cite{Hsu:1992tg}, employing a different technique and keeping terms to first order in $v$ in the non-relativistic limit. All these calculations assumed that the neutrino is massless and that there is only one flavor of neutrinos.

Despite being a very small effect, in Ref.~\cite{Fischbach:1996qf}, Fischbach claimed that if neutrinos were massless, the two-neutrino force between neutrons in a neutron star could raise the self energy of the system to a value that is much higher than the mass of the star itself. Without any other mechanism to stop this, Fischbach proposed that the neutrino is, in fact, massive. A massive mediator would shorten the range of the two-neutrino force and solve the problem. However, Smirnov and Vissani \cite{Smirnov:1996vj} posited that low-energy neutrinos created and subsequently captured in the star (the phenomenon is described in \cite{Smith:1985mta}) fill a degenerate Fermi sea that blocks the free propagation of the neutrinos that are responsible for the neutrino force. In response, Fischbach in Ref. \cite{Fischbach:1996ep} stated that more work needs to be done to understand the capturing process and that, for low energies, the two-neutrino force can be repulsive leading to the neutron star actually repelling neutrinos instead of filling up the Fermi sea. Then, Kiers and Tytgat in Ref.~\cite{Kiers:1997ty} argued that the neutrino self-energy does not destabilize the neutron star. Yet in a recent paper by Fischbach \cite{Fischbach:2018rfp}, he does not agree with that conclusion. In our work, we do not investigate this issue, and do not put any bound on the neutrino mass from neutron star considerations. Our focus is on aspects of the neutrino force that are relevant to atomic physics.

Following Fischbach's calculation of the potential due to massive Dirac neutrinos, Grifols et al.  \cite{Grifols:1996fk} calculated the same potential for massive Majorana neutrinos, which differ from Dirac neutrinos in the non-relativistic limit because of the different spinor structure of Majorana fermions. Their approach is the same as that in \cite{Feinberg:1968zz}. For future reference, the parity-conserving form of the two-neutrino potential to leading order in $v$ for the case of a single flavor of neutrinos with mass $m_{\nu}$ is given by
\beq
V_{\nu \nu}^{\text{Dirac}}(r) = \frac{G_F ^2 m_{\nu}^3}{4\pi^3 }\frac{K_3 (2m_{\nu}r)}{r^2},\qquad 
V_{\nu \nu}^{\text{Majorana}}(r) = \frac{G_F ^2 m_{\nu}^2}{2\pi^3 }\frac{K_2 (2m_{\nu}r)}{r^3},\label{Massive_nu}
\eeq 
where $K_n(x)$ is the $n$th order modified Bessel functions of the second kind.

An additional effect in neutrino physics, due to the non-zero masses, is flavor mixing (for a review, see, for example, Ref.~\cite{Strumia:2006db}). This phenomenon was incorporated into the computation of the two-neutrino force in Ref.~\cite{Lusignoli:2010gw}, although a closed form for the neutrino force was not attained. One can also look in \cite{Thien:2019ayp} for a treatment of the spin-independent part of the neutrino force with flavor mixing. Lastly, thermal corrections to the neutrino force, in both the Dirac and Majorana cases, were computed in \cite{Ferrer:1998ju}.

All the calculations mentioned above compute terms in the potential that are parity conserving, i.e. parity-violating terms have been ignored. In this work, we go beyond the leading-order results in $v$ and compute terms in the potential that are spin and momentum dependent and also parity violating. Our key results are described in section \ref{sec:pncforce}, and their implications are described in Sec.~\ref{sec:effects}. We keep terms to first order in $v$ in our non-relativistic calculation. 

\section{Observing Atomic Parity Violation -- a review}\label{sec:APV_review}

In this section, we review the concepts of Atomic Parity Violation (APV) that are relevant to the present work. We look at atomic parity violation from the perspective of transitions in atoms, more specifically, stimulated emission processes, wherein an emission is caused by shining light on a sample of atoms. For a more detailed review of APV from both theoretical and experimental perspectives, see Refs.~\cite{Bouchiat:1997mj,Guena:2005uj,Khriplovich:1991mba,Bernabeu:1974jt}.

The key idea behind looking for APV is to exploit the fact that in the presence of a parity violating term in the atomic Hamiltonian, the energy eigenstates have no definite parity. As per the well-known selection rules, electric dipole ($E1$) transitions happen between states of opposite parity while magnetic dipole ($M1$) transitions take place only between states of same parity. If the energy eigenstates, however, have no definite parity, then both $E1$ and $M1$ transitions are allowed between them. Since the parity violating interactions are usually very weak compared to the parity conserving ones, we treat them as perturbations to a parity conserving Hamiltonian.  Eigenstates of the full Hamiltonian, therefore, are superpositions of a predominant state of definite parity with small opposite parity corrections.

A direct consequence of the presence of parity-violating interactions is that left-polarized light has a different refractive index from right-polarized light in a sample of atomic vapors, which leads to optical rotation of light in the sample. This is the property that has been exploited to probe APV so far. An intuitive physical interpretation of this effect is due to Khriplovich~\cite{Khriplovich:1991mba}: Mixing opposite parity states in the hydrogen atom, for instance, results in the creation of a state wherein the electron effectively has a position-dependent spin orientation that assumes a helical shape. Recall that helical shapes of molecules lead to rotation of the plane of polarization of incident light on a sample. Classically speaking, this is because the electric field of light moving perpendicular to the helical axis causes electrons to produce an electric field along the helical axis, which induces a changing magnetic field perpendicular to the electric field. The combined effect of this is to rotate the plane of polarization of the incident electromagnetic wave.

\begin{figure}[t!]
\centerline{\includegraphics[width=4in]{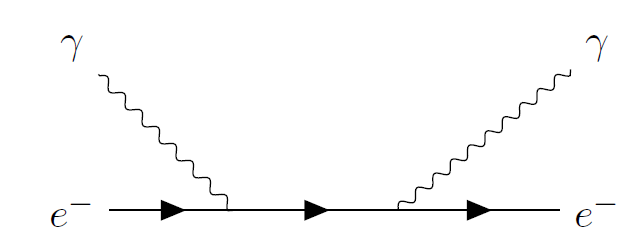}}
%
\caption{Stimulated emission as electron photon scattering}
\label{fig:stimemit}
\end{figure}

A stimulated emission transition is basically an electron-photon scattering process, represented by the diagram in Fig.~\ref{fig:stimemit}.
If both photons have the same polarization, and the photon is incident on a sample with electron density $N_e$, the scattering process can be translated into an index of refraction~\cite{Jackson:1998nia}. The refractive index $n_P$ depends on the polarization of the photon, labeled by the subscript $P=L,R$, and it is given by
\beq 
n_P^2(k) =  1+ \frac{4\pi N_e}{k^2}f_P(0). \label{refrindx}
\eeq
Here, $f_P(0)$ is the forward scattering amplitude for a photon with polarization $P$, and $k$ is the magnitude of the momentum of the photon.

\begin{figure}[t!]
\centerline{\includegraphics[width=4in]{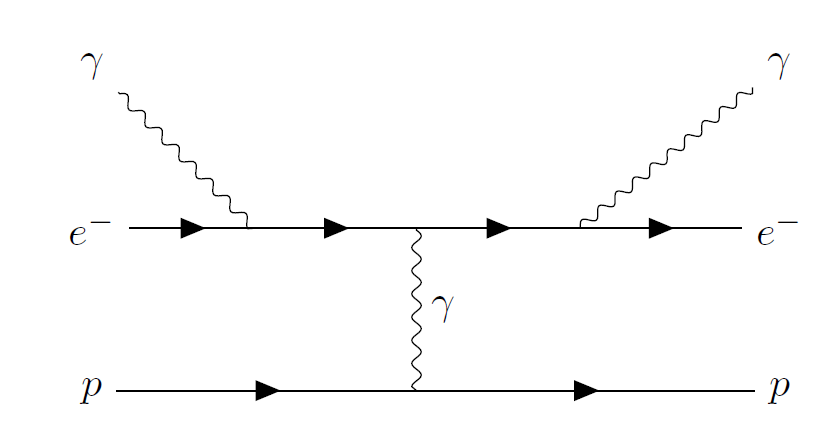}}
%
 \caption{Stimulated emission in hydrogen atom. The electron is shown to be bound to the proton by the mediation of a photon. This is the lowest order diagram at tree level.}
\label{fig:stimemitph}
\end{figure}

When the electron is bound in the electromagnetic field of a proton, as in hydrogen, the stimulated emission process, in the presence of Coulombic binding, is represented by the diagram in Fig.~\ref{fig:stimemitph}. We treat the proton as an elementary particle, since we work at energy scales small enough that the internal substructure of the proton can be ignored. In Fig.~\ref{fig:stimemit}, the proton can be seen as a correction to the electron propagator. Therefore, instead of calculating the transition amplitude using the matrix element from Feynman rules, we can alternatively first compute the static potential that mimics the scattering of the electron off the proton (in this case, the binding). This gives us, at lowest order, the Coulomb force. Thereafter, the external photons effectively become electromagnetic perturbations to the Coulomb field. We can now use time-dependent perturbation theory to calculate the transition amplitude. This is a simple quantum mechanical picture \cite{Emmons:1987jna} as opposed to a field theoretic perspective. In this picture, we usually talk about electric and magnetic dipole transitions whereas from the perspective of field theory, both transitions are just electron-photon scattering processes.

For incoming and outgoing photons with equal polarization, we can compute the refractive index in hydrogen gas using Eq.~\eqref{refrindx}. Note that parity is a good symmetry of QED, and hence $f_R(0)=f_L(0)$ for the process in Fig.~\ref{fig:stimemitph}. This implies that the refractive index is the same for left-handed and right-handed polarized photons. When parity is violated, the amplitudes for an incoming right circularly polarized photon and a left circularly polarized photon are different, that is $f_R(0)\neq f_L(0)$, hence $n_R(k)\neq n_L(k),$ causing optical rotation. In the SM the leading-order effect that violates parity is due to $Z$ exchange, and it arises from a diagram similar to the one in Fig.~\ref{fig:stimemitph} with the photon propagator replaced by a $Z$ propagator. We discuss this process in the next section.

\begin{figure}[t!]
  \centerline{\includegraphics[width=4in]{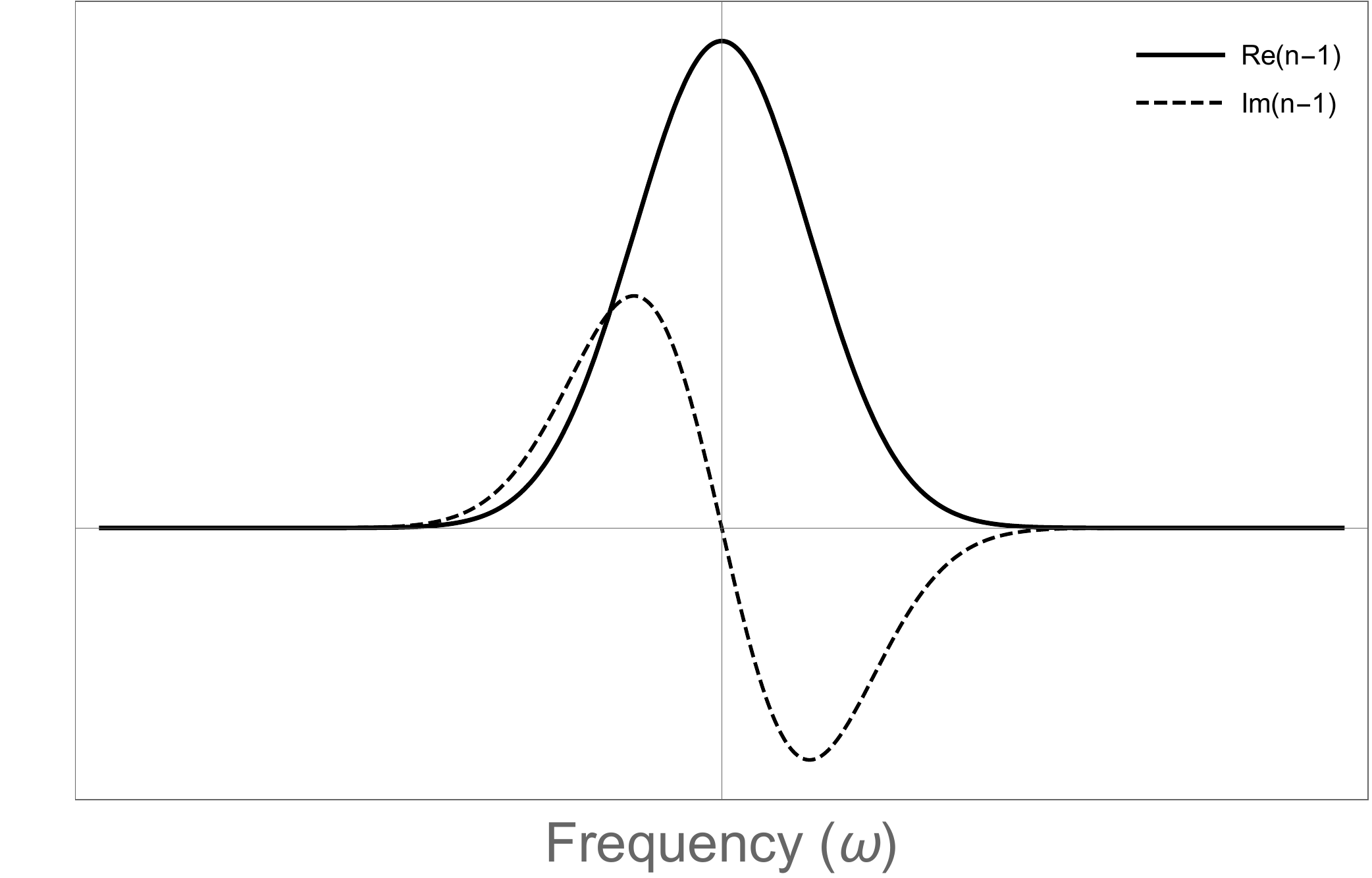}}
  \caption{The real and imaginary parts of the refractive index $n$ near a resonance. Absorption follows the imaginary part, while dispersion, and hence, optical rotation follows the real part.}
  \label{fig:KK}
\end{figure}

The refractive index, which we denote here by $n(\omega)$, of any material in general, and a gas of atoms in particular, has both real and imaginary components, corresponding to the dispersive and absorptive powers of the gas, respectively. The imaginary component is negligible for most values of the frequency, but it is large near bound-state resonances (i.e, when the energy of the incident photon equals the energy difference between two energy eigenstates), which is when the material becomes strongly absorbent. The real part is the well-known index of refraction. The Kramers-Kronig equations (see Ref. \cite{Toll:1956cya}) relate the two quantities as shown below:
\beq
\Re[n(\omega)] = 1 + \frac{2}{\pi}\int_{0}^{\infty}\D\omega'\  \frac{\omega'\ \Im[n(\omega')]}{\omega'^2-\omega^2}\label{refinKK}.
\eeq
Eq.~\eqref{refinKK} implies that the real part of the refractive index has a maximum near the resonance frequency and thus the local maxima of the real and imaginary parts are close in frequency, see Fig.~\ref{fig:KK}.\\

In a sample, the rotation of the plane of polarization of incident light is proportional to the real part of the refractive index \cite{hecht2015optics}:
\begin{equation} \label{eq:pre_angle_rot}
    \Phi =\frac{\pi L}{\lambda}\Re(n_R(\lambda)-n_L(\lambda)) 
\end{equation}
where $\Phi$ is the angle of rotation of the plane of polarization of incident light, $L$ is the length of the path of light through the sample and $\lambda$ is the wavelength of incident light.
Therefore, near a resonance, there is an enhancement of optical rotation in a material or a gas. \\

In time-dependent perturbation theory, one can compute the left-right asymmetry between the dipole-transition amplitudes (both electric and magnetic) for right-polarized and left-polarized light~\cite{Khriplovich:1991mba,Emmons:1987jna}. This asymmetry is related to the difference in the real part of the refractive indices for the two respective polarizations. Subsequent analysis yields $\Phi$, for states with the same predominant parity \cite{Emmons:1987jna} in terms of electric/magnetic dipole transition amplitudes. In the case that the wavelength is close to the difference in energy between two states of predominantly the same parity, the rotation is given by
\begin{equation} \label{eq:angle_rot}
    \Phi = \frac{4\pi L}{\lambda}\Re(n(\lambda)-1)R, \qquad  R = \Im\left(\frac{E1_{PV}}{M1}\right), 
\end{equation}
where $n(\lambda) = \frac{1}{2}\left(n_R(\lambda)+n_L(\lambda)\right)$ is the average refractive index of the sample, $E1_{PV}$ is the forbidden electric-dipole transition element, and $M1$ is the magnetic-dipole transition element between two states of the system with the same predominant parity.

A few points are in order regarding Eq.~\eqref{eq:angle_rot}:
\begin{enumerate}
    \item  Note that if parity is conserved, the $E1_{PV}$ amplitude is zero and hence the angle of rotation is zero. 
\item
One could also consider a situation where the two states are of opposite parity. In this case $M1=0$ and the effect is proportional to $M1_{PV}$ and we get a formula similar to that of Eq.~\eqref{eq:angle_rot}. Magnetic-dipole amplitudes, however, are much smaller than electric dipole amplitudes, so probing parity violating effects by observing parity-forbidden magnetic transitions is generally harder.
\item 
To obtain the largest angle of rotation, the wavelength $\lambda$ must be close to the energy spacing between the states that we are interested in, but far away enough to avoid resonance, as it is clear from Fig. \ref{fig:KK}. In other words, if $\omega_r$ is the frequency at which a resonance occurs, and $\omega$ is the frequency of the incident light, then for a large enough effect, we need to have $|\omega-\omega_r| \sim \Gamma$, where $\Gamma$ is the width of the resonance.
\end{enumerate}

In summary, an important consequence of APV is that, near a resonance, the emitted light has a rotated plane of polarization relative to the incident light. Experimentally, therefore, a measurement of this rotation is a measure of APV. From our theoretical perspective, the important quantity that encodes the effects of APV is $R$, defined in Eq.~(\ref{eq:angle_rot}).

\section{Parity violating forces in atomic systems}\label{sec:pncforce}

\subsection{Generic effects}
The general expression for a non-relativistic potential between two fermions contains only a handful of terms -- the only difference between the potentials mediated by different mechanisms is in the numerical coefficients coming with each term and the form of the radial function \cite{Dobrescu:2006au}.\\

Consider a generic atom with a nucleon of mass $m_N$. We are looking for the parity violating potential due to some Feynman diagram. To that end, we make two simplifying assumptions:
\begin{enumerate}
\item
We consider a static nucleus, that is, we neglect effects that scale like $m_e/m_N$.
\item
We treat the electron velocity, $v_e$, as a small parameter and keep only terms linear in $v_e$.
\end{enumerate}
Under these assumptions, the most general form of the parity-violating potential from \cite{Dobrescu:2006au} reduces to the following:
\begin{equation}
   V_{PNC}(r)= H_1 F(r)\vec{\sigma}_e \cdot\vec{v}_e + H_2 F(r)\vec{\sigma}_N\cdot\vec{v}_e+C(\vec{\sigma}_e\times \vec{\sigma}_N)\cdot \vec{\nabla}\left[F(r)\right],\label{treepotgen}
\end{equation}
where $\vec{\sigma}_e/2$ is the spin of the electron, $\vec{\sigma}_N/2$ is the net nuclear spin, $H_1$, $H_2$ (for ``helicity'', since the corresponding terms look like helicity) and $C$ (for cross-product) are real constants, and $F(r)$ is a radial real function. 

The values of the $H_1$, $H_2$, $C$, and $F(r)$ depend on the specific diagram. In case there are several diagrams, each diagram contributes linearly to the total potential, so we can write 
\beq
V_{PNC}(r)=\sum_i V_{PNC}^i(r) \label{eq:vpncsum}
\eeq
and we add a sub-index $i$ to $H_1$, $H_2$, $C$, and $F(r)$.

In the following sections, we shall consider the special case of the hydrogen atom. While experiments are not done with it, it simplifies the theoretical investigation. When we consider hydrogen, we replace the sub-index $N$ with $p$.

\subsection{The tree-level process}\label{treeprocess}

\begin{figure}[t!]
\centerline{\includegraphics[width=4in]{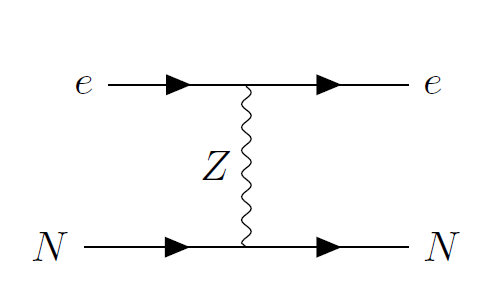}}
%
\caption{Tree-level interaction between the electron and a nucleus.} \label{fig:NeZNe}
\end{figure}

We begin by briefly revisiting the effective parity-violating potential due to the interaction between an electron and a nucleus at tree level via $Z$ exchange in the SM as depicted in Fig.~\ref{fig:NeZNe}. 
In the SM, the coupling of the $Z$ boson to a pair of identical fermions is given by 
\beq
\mathcal{L}_{Z\bar{\psi}{\psi}} = \frac{1}{2}\frac{g}{\cos \theta _W}\bar{\psi}\left[ (g_V^\psi-g_A^\psi\gfive)\slashed{Z}\psi \right],
\eeq 
where $\theta_W$ is the Weak angle. $g^{\psi}_V$ and $g^{\psi}_A$ are the vectorial and axial couplings of the fermion $\psi$ to the $Z$ boson.
As an example, the coupling constants  for the electron and the proton (which can be treated as an elementary particle at energy scales relevant to atomic physics) are given by:
\begin{equation}
g_V^e \,=\, \left(-\frac{1}{2}+ 2 \sin ^2 \theta_W \right),\quad g_A^e\, =\, -\frac{1}{2}, \quad
g_V^p\,= \,\left(\frac{1}{2}- 2 \sin ^2 \theta_W \right), \quad  g_A^p\,=\, \frac{G_A}{2}, \label{eq:defGa}
\end{equation}
where $G_A\approx 1.25$ \cite{Mund:2012fq} is the axial form factor of the proton.

The resulting parity-violating potential is given by Eq. \eqref{treepotgen} with the constants and the radial function given by:

\begin{eqnarray}
H_1 &=&  H_1^{\text{tree}}=  \frac{g^2}{2 \cos^2 \theta_W}g_A^e g_V^p , \label{eq:treeh1}\\
H_2 &=& H_2^{\text{tree}} = \frac{g^2}{2 \cos^2 \theta_W}g_V^e g_A^p ,\\
C &=& C^{\text{tree}}= \frac{g^2}{2 \cos^2 \theta_W}\frac{g_V^e g_A^p}{2 m_e},\\
F(r) &=& F^{\text{tree}}(r)= \frac{e^{-m_Z r}}{4\pi r}.\label{eq:treef1}
\end{eqnarray}

In the APV literature, most notably in \cite{Bouchiat:1974zz}, the terms that depend on nuclear spin (that is, terms that come with $H_2$ and $C$) are ignored. This is because, in most heavy atoms used in APV experiments, the nuclei have paired nucleons with opposite spins, and a net nuclear spin of zero. Thus, terms in the potential containing the nuclear spin vanish. This is not true for the case of hydrogen, where the nucleus consists of just one spin-half proton.

\subsection{Loop level processes: The effective four-Fermi operator with neutrinos}
\begin{figure}[t!]
\subfloat[]{
\includegraphics[width=2in]{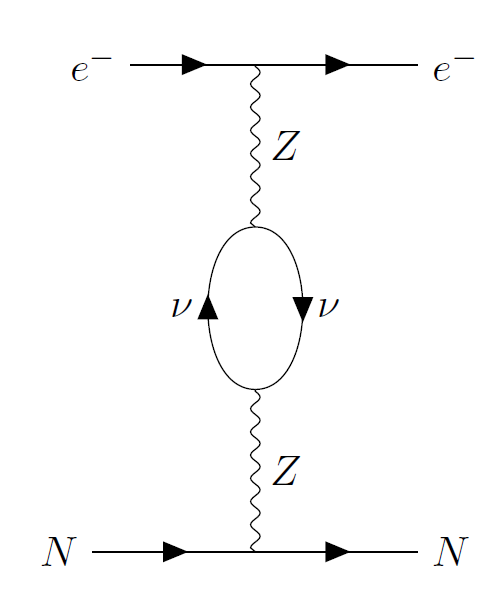}
%
\label{fig:loopZ}}
\subfloat[]{
\includegraphics[width=2in]{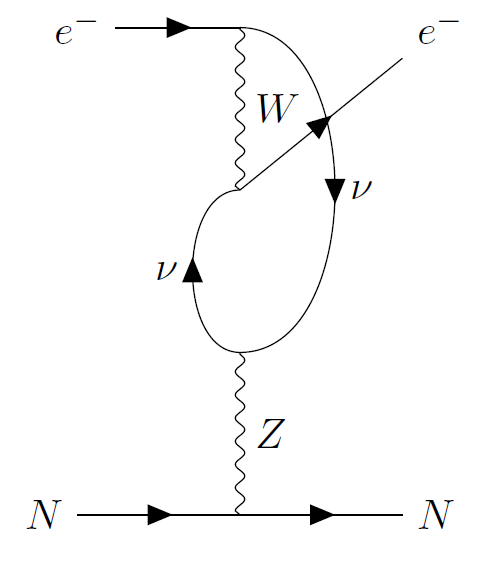}
%
\label{fig:loopW}}
\caption{The loop level diagrams that contribute to the binding of the electron to the nucleus in an atomic system.}
\label{fig:loopdiag}
\end{figure}

Now that we have discussed the tree level diagram that violates parity, we move on to loop level effects. The diagrams that contribute to atomic parity violation at one loop are given in Fig.~\ref{fig:loopdiag}. At atomic energy scales, the use  of the four-Fermi approximation is well justified and so in this section, we will derive expressions for the four-Fermi vertices with two fermions of the same type $\psi$ and two neutrinos. 

\begin{figure}[t!]
\subfloat[]{
\includegraphics[width=2in]{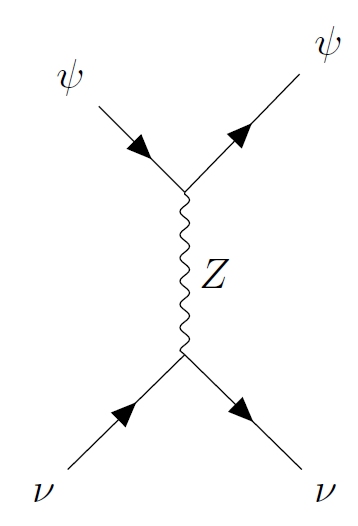}
%
%
\label{fig:OZ}}
\subfloat[]{
\includegraphics[width=2in]{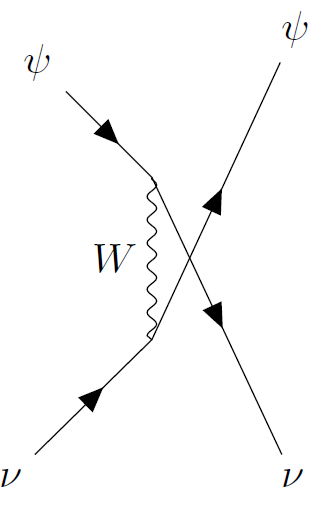}
%
%
\label{fig:OW}}
\caption{The two diagrams that contribute to the effective four-Fermi vertex for two neutrinos and two fermions $\psi$. The $Z$-diagram in Fig. \ref{fig:OZ} corresponds to the effective operator $\mathcal{O}_Z$. The $W$ diagram in Fig. \ref{fig:OW} corresponds to the effective operator $\mathcal{O}_W$.}
\label{fig:fourfermi}
\end{figure}

In the SM, the four-Fermi interactions between two neutrinos and two fermions are obtained by integrating out the $Z$ and $W$ bosons in the diagrams shown in Fig.~\ref{fig:fourfermi}. However, since we consider massive neutrinos, we need to incorporate flavor mixing. The $Z$-boson case is simple because the interactions of neutrinos with the $Z$ boson is universal and thus diagonal in any basis:
\begin{equation} \label{eq:L_neu_Z}
\mathcal{L}_{Z}= -\frac{g}{2c_W}\delta_{ij}\bar{\nu}_i\slashed{Z}\nu_j ,
\end{equation}
with $c_W \equiv \cos\theta_W$. The corresponding four-Fermi operator for a vertex involving two fermions $\psi$, and two neutrino mass eigenstates, $\nu_i$  and $\nu_j$, due to $Z$ exchange is therefore
\beq \label{eq:OZ}
(\mathcal{O}_Z)_{ij}  = -\frac{g^2}{8m_Z ^2 c_W^2}[\bar{\psi}\gmu (g^{\psi}_V - g_A^{\psi} \gfive)\psi]\delta_{ij}[\bar{\nu}_j\gMu(1-\gfive)\nu_i],
\eeq
where $g_A^\psi$ and $g_V^\psi$ are defined above Eq.~\eqref{eq:defGa}. 

The case of the $W$ exchange is more complicated as we need to take into account the non-diagonal nature of the flavor mixing. The $W$ interaction Lagrangian in the mass basis for the neutrinos is given by:
\begin{equation} \label{eq:L_neu_flavor}
\mathcal{L}_{W}=  - \frac{g}{\sqrt{2}}U_{\alpha i} \bar{\ell}_{L\alpha} \slashed{W}\nu_i,
\end{equation}
where  the fields $\ell$ represent leptons and $i$ ($\alpha)$ represents mass (flavor) indices, and $U_{\alpha i}$ are the elements of the Pontecorvo-Maki-Nakagawa-Sakata (PMNS) matrix.
The operator for the case of two external $\psi$ leptons of flavor $\alpha$ and two neutrino mass eigenstates $i$ and $j$ is then given by
\begin{eqnarray} \label{eq:OW}
(\mathcal{O}_W)_{ij} &=& -\frac{g^2}{8m_W ^2}U_{\alpha j}U^*_{\alpha i}[\bar{\nu}_j\gmu (1 - \gfive)\psi][\bar{\psi}\gMu(1-\gfive)\nu_i], \nonumber \\
&=& -\frac{g^2}{8m_W ^2}U_{\alpha j}U^*_{\alpha i}[\bar{\psi}\gmu (1 - \gfive)\psi][\bar{\nu_j}\gMu(1-\gfive)\nu_i], \label{fourfermi}
\end{eqnarray}
where we used the Fierz transformations to obtain the second line.

The sum of the operators in Eqs.~\eqref{eq:OZ} and \eqref{eq:OW} yields the four-fermion vertex between two neutrino mass eigenstates and two $\psi$ leptons. Using $G_F = g^2/4\sqrt{2}m_W^2$, we obtain
\begin{eqnarray}
\mathcal{O}_{ij} &=& (\mathcal{O}_Z)_{ij}+(\mathcal{O}_W)_{ij}\\
&=&  -\frac{G_F}{\sqrt{2}}\left[\bar{\psi}\gmu \{\delta_{ij}(g_V^{\psi} - g^{\psi}_A \gfive)+U_{\alpha j}U_{\alpha i}^*(1-\gfive)\}\psi\right] \left[\bar{\nu}_j \gMu(1-\gfive) \nu_i\right], \nonumber \\
&=& -\frac{G_F}{\sqrt{2}}\left[\bar{\psi}\gmu (a^{\psi}_{ij}-b^{\psi}_{ij}\gfive)\psi\right] \left[\bar{\nu}_j \gMu(1-\gfive)\nu_i\right]. \nonumber \label{eq:effcoup}
\end{eqnarray}
We emphasize that there is no sum over $i$, $j$ or $\alpha$ here. In Eq.~\eqref{eq:effcoup}, we introduced the effective vectorial and axial couplings, $a_{ij}$ and $b_{ij}$ respectively, in terms of the couplings to the $Z$. If $\psi$ is a lepton and therefore has a flavor index $\alpha$, we have:
\begin{eqnarray}
a^{\psi}_{ij} = \delta_{ij} g^{\psi}_V+U_{\alpha j}U_{\alpha i}^* , \quad
b^{\psi}_{ij} = \delta_{ij} g^{\psi}_A+U_{\alpha j}U_{\alpha i}^* .\label{eq:abijl}
\end{eqnarray}
If $\psi$ were not a lepton, it would not couple to neutrinos through the $W$, and therefore the PMNS matrix would not be involved. Then we would have:
\begin{eqnarray}
a^{\psi}_{ij} = \delta_{ij} g^{\psi}_V, \quad b^{\psi}_{ij} = \delta_{ij} g^{\psi}_A, \label{eq:abijn}
\end{eqnarray} 

In order to compute the neutrino force between two fermionic species $\psi_1$ and $\psi_2$, we need to insert the operator $\mathcal{O}_{ij}$ twice in order to obtain the diagram in Fig.~\ref{fig:2nu}. If both $\psi_1$ and $\psi_2$ are leptons, we have nine diagrams from assigning three neutrino mass eigenstates into the two propagators. Each diagram is labeled by two indices $i$ and $j$, and we sum over them. If $\psi_1$ or $\psi_2$ is a non-lepton, then the only possible four-Fermi vertices are the ones with both neutrinos in the same mass eigenstate. Thus, there are three diagrams over which to sum over. We only need one label $i=1,2,3$ to denote a diagram since the effective couplings $a$ and $b$ are diagonal. We shall make use of precisely this fact to explore APV in the simplest atomic system, i.e, the hydrogen atom, in Sec.~\ref{neutrinoH}.

\subsection{The photon penguin} \label{sec:photon-pen}
\begin{figure}[t!]
\subfloat[]{
\includegraphics[width=4in]{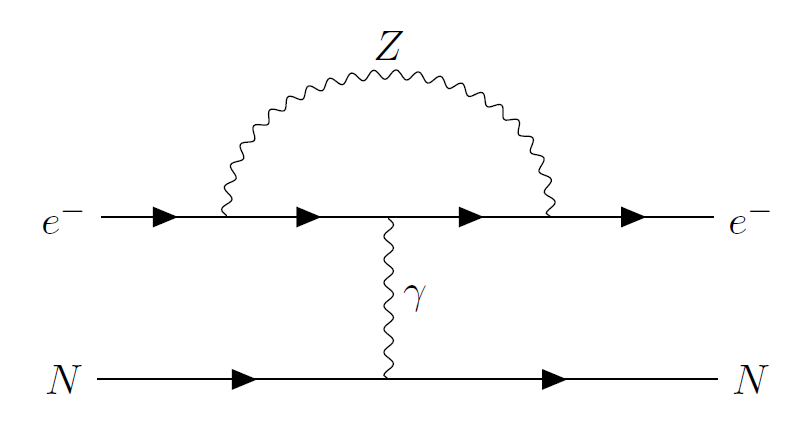}
%
\label{fig:pp}}
\subfloat[]{
\includegraphics[width=2in]{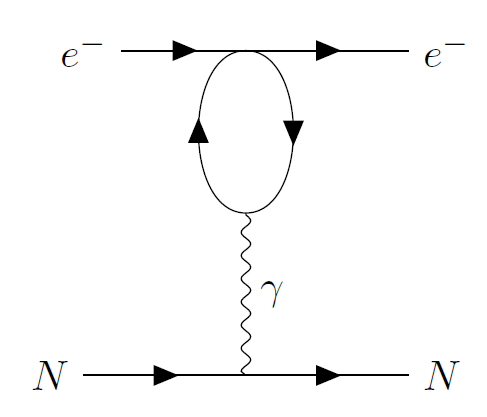}
%
\label{fig:ppz}}
\caption{The photon penguin is shown in Fig. \ref{fig:pp}, and again in Fig. \ref{fig:ppz} after integrating out the $Z$ boson.}
\label{fig:PP}
\end{figure}

In this subsection, we digress to talk about another possible parity violating diagram in our atomic system, the photon penguin, shown in Fig.~\ref{fig:pp}. This diagram is also parity violating since it has two weak interaction vertices. However, it does not give rise to a long ranged parity violating potential, as we discuss below.

Assuming that the momentum transfer is much smaller than the $Z$ boson mass, we can integrate out the $Z$ resulting in an effective photon penguin diagram as shown in Fig.~\ref{fig:ppz}. Using the same approach as in Ref.~\cite{Feinberg:1989ps}, the parity-violating potential is found to be in accordance with Eq.~\eqref{treepotgen}, with the quantities $H_1$, $H_2$, $C$, and $F(r)$ given by:
\begin{eqnarray}
H_1 &=&  H_1^{\text{penguin}}=  (g_V ^e)(g_A ^e) G_F \ \alpha \ m_e, \label{eq:pengh1}\\
H_2 &=& H_2^{\text{penguin}} = 0 ,\\
C &=& C^{\text{penguin}}= 0,\\
F(r) &=& F^{\text{penguin}}(r)\label{eq:pengf1}
\end{eqnarray}
where the explicit calculation gives
\beq
F^{\text{penguin}}(r) = 2 e^{-2m_e r} \int_0^{\infty} dx \  e^{-rx} \frac{\sqrt{x^2+ 4x m_e}}{(x+2m_e)^2} \left( x^2 + 4x m_e +6m_e^2 \right).
\eeq

We were unable to find a closed form of $F^{\text{penguin}}(r)$. However, for our purpose, all we care about is the $r \gg m_e$ limit. In that limit,
we obtain the asymptotic form of $F^{\text{penguin}}(r)$ to be:
\beq \label{eq:F-large-r}
F^{\text{penguin}}(r)\sim 12 \Gamma \left(\frac{3}{2}\right) \sqrt{m_e} \frac{e^{-2m_e r}}{r^{5/2}} .
\eeq
The main conclusion from Eq.~(\ref{eq:F-large-r}) is that 
the potential due to the photon penguin far away from the nucleus is suppressed by an exponential term that has a range given by $(2m_e)^{-1}$. The factor of 2 is because the penguin diagram consists effectively of two exchanged electrons, just like in the neutrino case where we exchanged two neutrinos. The potential is sensitive to the mass of the electron because there is a branch cut in the scattering amplitude for the diagram in the complex$-t$ plane starting at $t = 4m_e^2$.

Some important differences between this potential and the potential with two internal neutrinos are as follows:
\begin{enumerate}
\item The potential due to the photon penguin has only one factor of $G_F$, as opposed to the neutrino case which is suppressed by two factors of $G_F$. Yet, given the short range nature of the photon penguin for atomic physics, which is what our concern in this work, this force is not relevant. 
\item 
The other possible photon penguin has a $Z$ boson joining the proton legs. But as it is clear from this calculation, the diagram analogous to Fig.~\ref{fig:ppz} for this process has a proton in the loop, and thus the radial function will have its range governed by the proton mass. This makes the parity violating force shorter ranged than the force obtained from the photon penguin with the electron loop, and thus negligible.
\end{enumerate}
To conclude, the key takeaway is that the photon penguin does not contribute to long rage parity violation at atomic scales. We can safely ignore its effects for the systems that we care about in this paper.

\section{The neutrino force in the hydrogen atom}\label{neutrinoH}

We now apply the results obtained above to the hydrogen atom. In the hydrogen atom, the proton does not couple to the neutrinos through the $W$ boson, and so the only diagrams that contribute are the three diagrams with the same neutrino  mass eigenstate on both propagators in the loop.
Using Eqs.~\eqref{eq:abijl} and \eqref{eq:abijn}, we find that in this case, the corresponding couplings are diagonal and are given by (superscripts refer to the electron and the proton respectively)
\begin{eqnarray}
a^e_{ii} &=& \left(-\frac{1}{2}+ 2 s_W^2+ |U_{ei}|^2 \right), \quad a^p_{ii} = \left(\frac{1}{2}- 2 s_W^2 \right)\nonumber \\
b^e_{ii} &=& \left(-\frac{1}{2}+|U_{ei}|^2 \right), \quad b^p_{ii} = \frac{G_A}{2} \approx 0.625, \label{Eq:couplingsmix}
\end{eqnarray}
where $G_A$ is the axial form factor, as defined below Eq.~\eqref{eq:defGa}, and $s_W = \sin\theta_W$. Since both propagators have the same mass eigenstate, the non-diagonal entries in $a_{ij}$ and $b_{ij}$ are zero. For the same reason, we only keep one index $i$ from now on. 

Using the couplings from Eq.~\eqref{Eq:couplingsmix}, we calculate the parity-violating potential from the neutrino loop, which results in a form given by Eq.~\eqref{eq:vpncsum} (see appendix \ref{app:pnc} for details of the calculation).
with the constants and the radial function given by (no sum over $i$ in any of the expressions):
\begin{eqnarray}
 H_{1i} &=& H_{1i}^{\text{loop}}= - 2\frac{a^p_{i}b^e_{i}}{m_e}   ,\label{eq:hcf1} \\
 H_{2i} &=& H_{2i}^{\text{loop}} = 2\frac{a^e_{i}b^p_{i}}{m_e},\label{eq:hcf2}\\
 C_i &=& C_i^{\text{loop}}= \left( \frac{a^e_{i}b^p_{i}}{m_e}+ \frac{a^p_{i}b^e_{i} }{m_p}\right),\label{eq:hcf3}\\
 F_i &=& F_i^{\text{loop}}(r)= V_{\nu_i \nu_i}(r)\label{eq:hcf4} ,
\end{eqnarray}
where  $V_{\nu_i \nu_i}(r)$ can be found in Eq. \eqref{Massive_nu}. 

Using the fact that $s_W^2 \approx 0.23$, so that $a^p_{i}$ is very small and that $m_e \ll m_p$, we note that $H_{1i}$ is negligible. The parity-violating potential then simplifies to:
\begin{eqnarray}
V_{PNC}^{\text{loop}} \approx \sum_i \frac{G_A}{m_e}\left(-\frac{1}{4}+  s_W^2 +\frac{1}{2}|U_{ei}|^2\right)\left[(2\vec{\sigma}_p\cdot\vec{p_e})V_{\nu_i \nu_i}(r) +(\vec{\sigma_e} \times \vec{\sigma_p})\cdot\vec{\nabla}V_{\nu_i \nu_i} (r)\right]. \label{vpncloop}
\end{eqnarray}
Eqs.~\eqref{eq:hcf1}-\eqref{vpncloop} are the key results in our work. The parity-violating terms obtained here have the same spin structure as in the case of the tree-level potential, but the radial behavior is different. Investigation of these terms in the neutrino potential has not been carried out before.

\section{Effects of the neutrino force on hydrogen eigenstates and transitions}\label{sec:effects}

In this section, we treat the neutrino potential in Eq.~\eqref{vpncloop} as a perturbation to the hydrogen atom Hamiltonian. We work in the limit $m_p \to \infty$, so that the proton is essentially static. We assume that the neutrino is of Dirac nature subsequently in this paper, but one could also treat them as Majorana fermions and perform an analogous computation. 

The neutrino force is much smaller than the fine or hyperfine interactions and therefore, we need to include the fine-structure and the hyperfine splittings as well in our calculations. As always, we should look for an operator that commutes with the neutrino potential, and use the eigenbasis of this operator as the basis of choice in first-order degenerate perturbation theory. Since the neutrino potential is a scalar, we know that an operator that commutes with it is $\hat{F}^2$, where $$\vec{F} \equiv \vec{L}_e+\vec{S}_e+\vec{S}_p$$ is the total angular momentum of the entire system. We also define $\vec{J}\equiv \vec{L}_e+\vec{S}_e$ as the total angular momenta of the electron alone.

The unperturbed eigenstates $\ket{n,f,m_f,j,\ell,s_p,s_e}$ with which we work are simultaneous eigenstates of $\hat{H}_0,\hat{F}^2,\hat{F}_z,\hat{J}^2,\hat{L}_e^2,\hat{S}_p^2$ and $\hat{S}_e^2$, where  $\hat{H}_0 = \vec{p}^2/2m_e - e^2/r$ is the unperturbed hydrogen atom with only the Coulombic interaction. The eigenvalues of $\hat{F}^2,\hat{F}_z,\hat{J}^2,\hat{L}_e^2,\hat{S}_p^2$ and $\hat{S}_e^2$ are $f(f+1),m_f,j(j+1),\ell(\ell+1),s_p(s_p+1)$ and $s_e(s_e+1)$ respectively. Every state is thus described by 7 quantum numbers. But $s_e=s_p = 1/2$ are fixed numbers, and so we really need just 5 numbers to label a state. This is indeed what we expect since the hydrogen atom has a total of 8 degrees of freedom (dof): there are 3 position dof and 1 spin dof each for the electron and the proton. However, we do not care about the three dof of the center of mass, leaving us with 5 dof to describe the internal dynamics of our system. 

The angular momentum states can be constructed using the standard procedure of angular momentum addition using Clebsch-Gordon coefficients, as done in Ref.~\cite{shankar2012principles}, for instance. The orbital angular momentum of the electron $\ell$ takes values $0,1,2,\dots$ Depending on $\ell$, the result of the angular-momentum addition of one orbital angular momentum and two spin-1/2 systems (the electron and the proton are both spin-1/2)  can be summarized in the following notation:
     \begin{eqnarray}
    \ell\otimes\frac{1}{2}\otimes\frac{1}{2} = \underbrace{(\ell+1)\oplus \ell}_{j = (2\ell+1)/2}\oplus\underbrace{\ell\oplus (\ell-1)}_{j = (2\ell-1)/2}.
    \end{eqnarray}
These vector spaces contain eigenstates of the hydrogen atom written in the basis of $\hat{F}^2$ for a given principal quantum number $n$. The first two vector spaces in the direct sum consist of states with a well-defined value of $j = (2\ell+1)/2$, while the latter two vector spaces have well-defined $j = (2\ell-1)/2$. 

In the unperturbed hydrogen atom, all these states would be degenerate. But with the perturbations, such as the fine structure corrections and the hyperfine splitting interactions included, the degeneracy is lifted, and only the degeneracy in $m_f$ is left. The energy of an eigenstate with quantum numbers $f,j,\ell,s_e=s_p=\frac{1}{2}$, for the case where $\ell>0$, is given by (see Ref.~\cite{weissbluth2012atoms})
\begin{eqnarray}
E_{nfj\ell} &=& (E_0)_n+ (E_{\text{fine}})_{nj} + (E_{\text{hyperfine}})_{nfj\ell}  \label{Henergy}
\end{eqnarray}
where:
\begin{eqnarray}
  (E_0)_n &=& -\frac{\alpha^2 m_e}{2n^2}, \\
 (E_{\text{fine}})_{nj}&=& -\frac{\alpha^4 m_e}{2n^4}\left(\frac{n}{j+\frac{1}{2}}-\frac{3}{4}\right),\\
 (E_{\text{hyperfine}})_{nfj\ell}&=&\frac{ \alpha ^4 g_p}{m_p} a_0^3\frac{ \ell (\ell+1) m_e^2 \left(f (f+1)-j (j+1)-\frac{3}{4}\right)}{4 j (j+1) }\mean{\frac{1}{r^3}}_{n\ell}
\end{eqnarray}
are the energies contributed by the Coulombic potential, fine structure and hyperfine interactions respectively, $r$ is the radial coordinate of the electron, $a_0 = \left(m_e \alpha\right)^{-1}$ is the Bohr radius, and $g_p \approx 5.56$ is the $g$-factor of the proton~\cite{Mooser:2014vla}.

As a reminder, in first-order perturbation theory, in the presence of a perturbation $V$, the corrected states are given by
\begin{equation}
\ket{\psi_{q}^1} = \ket{\psi_{q}^0} + \sum_{p \neq q} \frac{\bra{\psi_p ^0}V\ket{\psi_q ^0}}{E_q ^0 - E_p ^0}\ket{\psi_p ^0} \label{pert}
\end{equation}
Here, $\ket{\psi_{p}^0}$ are the states in our chosen eigenbasis. Note that in this basis our perturbation is diagonal in each degenerate subspace. Under the perturbation, we say that the states in this basis ``mix'' among themselves to give the true eigenstates of the system.

The energy difference between states of different $n$ is much larger than that for those states with the same principal quantum number. Since the corrections to the eigenstates in perturbation theory go as $(\Delta E)^{-1}$, we keep corrections contributed by states with the same $n$ as our unperturbed states when calculating opposite-parity corrections to eigenstates in first-order perturbation theory.

Note that states mix among themselves under a scalar perturbation only when they have the same value of $f$. But, for any eigenstate of $\hat{F}^2$, the correcting states have a different value of $\ell$ if the perturbation violates parity. Therefore, under the effect of a parity-violating perturbation, a state attains an opposite parity admixture as expected. As discussed in Sec.~\ref{sec:APV_review}, both $E1$ and $M1$ transitions are therefore allowed between the actual eigenstates and we can expect to see an interference of $E1$ and $M1$ amplitudes that leads to optical rotation in a sample of atomic hydrogen. In Sec.~\ref{sec:sample_calc}, we shall compute this effect for certain states in hydrogen.

Parity violation in hydrogen is also manifest from the tree-level $Z$-potential. Intuitively, for states with $\ell=0$ , this tree-level process should completely overpower the neutrino loop diagram because these states have strong presence at the origin, which is also where the $Z$-potential has strong support. Thus, isolating an observable effect from the loop is unfeasible for such states. Higher-$\ell$ states do not have strong support at the origin and it would appear that the $Z$-potential does not have much effect on them. However, special care is needed, as we discuss in the next paragraph.

The neutrino-loop potential is highly singular. Therefore, at very short distances, the four-Fermi theory breaks down and we cannot trust our calculations all the way to $r=0$. (In order to still use our theory at short distances, we need to follow the methodology described in~\cite{Lepage:1997cs}. See also \cite{Frank:1971xx} for a discussion of singular potentials in the Schr\"{o}dinger equation. Alternatively, we could simply compute the diagrams in Fig.~\ref{fig:loopdiag} explicitly without integrating out the heavy bosons, as in the paper by Asaka et.al. \cite{Asaka:2018qfg}) However, if the momentum transfer is much smaller than the mass of the $Z$ boson or, in other words, the length scales are larger than $m_Z^{-1}$, then our calculations can still be trusted. Thus, we are interested in those high-enough values of $\ell$ for which the effects of the loop potential dominate over the $Z$-potential, while being far enough from the origin such that the four-Fermi theory is valid.
In the next two subsections, we select those eigenstates of hydrogen that are suitable for the task and show that, for states with orbital angular momentum $\ell\geq 2$, our conditions are met. A full computation of the loop diagrams as done in \cite{Asaka:2018qfg} would give us finite results for $\ell =0,1$, but is not necessary here since for $\ell <2$, the effects of the tree level $Z$-diagram dominates over the neutrino mediated diagrams that we are interested in.  We ultimately deal with eigenstates of $\hat{F}^2$, which do not have definite $\ell$, so we need to make sure that the eigenstate of $\hat{F}^2$ is a superposition of eigenstates of $\hat{L}^2$ with $\ell \geq 2$. 

\subsection{Matrix elements of the tree-level potential}
In order to extract some features of the tree-level parity violating potential, we write out the potential here as given in Eqs.~\eqref{eq:treeh1}-\eqref{eq:treef1}, but we suppress most of the dimensionless constants for the sake of clarity:
\begin{align}
V_{PNC}^{\text{tree}} \sim \frac{g^2}{m_e}\left[\frac{e^{-m_Z r}}{r}\vec{\sigma}_e\cdot\vec{p} +\frac{e^{-m_Z r}}{r}\vec{\sigma}_p\cdot\vec{p} + (\vec{\sigma}_e\times\vec{\sigma}_p)\cdot\vec{\nabla}\left(\frac{e^{-m_Z r}}{r}\right)\right] .  \label{vtree}
\end{align}
We are interested in computing the matrix elements of this potential in the space of hydrogen eigenfunctions. In this section, we simply consider the radial integrals in the matrix elements since the angular integrals simply give some $\mathcal{O}(1)$ number upon evaluation. We define $\eta \equiv r/a_0$, where $r$ is the radial coordinate. The radial part of the wavefunction, close to the origin, behaves as $u(\eta)\sim \eta^\ell$. Given this, we can write the matrix element as an integral:
\beq
\mel{V_{PNC}^{\text{tree}}}{n\ell m}{n'\ell' m'} \sim \int_0 ^{\infty} \D \eta  \ \eta^2\  \eta^{\ell'} V_{PNC}^{\text{tree}}(\eta) \eta^\ell,
\eeq
Note that, although the above dependence of the wavefunction is only correct near the origin, we integrate all the way to $\eta \to  \infty$ because the potential drops very rapidly in magnitude and so the contribution far away from zero from the wavefunction is negligible anyway.

Terms in the potential of Eq.~\eqref{vtree} that have angular dependence make the integral vanish unless $\ell'=\ell\pm 1$ (from the properties of the spherical harmonics). Without loss of generality, we take the smaller of the two to be $\ell$, and the larger to be $\ell+1$. Then the matrix element goes as (notice that the  momentum operator introduces a factor of $1/\eta$, as does a gradient)
\begin{eqnarray}
\mel{V_{PNC}^{\text{tree}}}{n\ell m}{n',\ell\pm 1,\, m'}  &\sim & \frac{\alpha}{m_e a_0^2}\int_0 ^{\infty}\D \eta \  \eta^{\ell+1} \exp\left(-m_Z a_0 \eta\right) \eta^\ell, \nonumber \\
&& \sim \frac{\alpha ^{2\ell +5} m_e^{2\ell +3}}{m_Z^{2\ell +2}}= m_e \alpha ^{2\ell+5} \left(\frac{m_e}{m_Z}\right)^{2\ell+2}\label{tree1}.
\end{eqnarray} 

\subsection{Matrix elements of the neutrino loop potential}

There are two terms in the loop potential~\eqref{vpncloop}: the ``helicity'' term and the spin-cross term. Once again, we consider only the radial integrals since the angular integrals give some $\mathcal{O}(1)$ number. The radial dependence of the integrands in the matrix elements is roughly the same, since the momentum operator and the gradient operator have the same radial structure.

The leading-order dependence of the parity non-conserving loop terms goes like $G_F ^2/m_e r^6$. Matrix elements for this operator go as
\begin{eqnarray}
\mel{V_{PNC}^{\text{loop}}}{n\ell m}{n'\ell' m'} &\sim \frac{G_F ^2}{m_e a_0^6}\int \D \eta \ \eta^2 \eta^{\ell'}\left(\frac{1}{\eta^6}\right)\eta^\ell \exp\left[-\eta\left(\frac{1}{n}+\frac{1}{n'}\right)\right] \nonumber \\
&\sim \frac{\alpha^2}{m_e m _Z ^4 a_0^6}\int \D \eta \ \eta^2 \eta^{\ell'}\left(\frac{1}{\eta^6}\right)\eta^\ell\exp\left[-\eta\left(\frac{1}{n}+\frac{1}{n'}\right)\right].
\end{eqnarray}
In the expression above, $\left(\frac{1}{n}+\frac{1}{n'}\right)\sim \mathcal{O}(1)$ number, which yields some exponential suppression. Let us denote this number by $n_{sup}$.
The angular integrals vanish unless $\ell'=\ell\pm 1$ and, like before, we can estimate a naive dependence of the wave function on $\alpha$, $m_e$, etc. We write
\begin{eqnarray}
\mel{V_{PNC}^{\text{loop}}}{n'(\ell+1)m'}{n\ell m} &\sim \frac{\alpha^2}{m_e m _Z ^4 a_0^6}\int \D \eta \ \eta^2 \eta^{\ell+1}\left(\frac{1}{r^6}\right)\eta^\ell\exp(-n_{sup} \eta) \nonumber\\
&\sim \frac{\alpha^2}{m_e m _Z ^4 a_0^6}\int \D \eta \ \eta ^{2\ell-3}\exp(-n_{sup} \eta).\label{eq:2l-3}
\end{eqnarray}
Now, we have the following sub-cases:
\begin{enumerate}
\item For $\ell=0$ and $\ell=1$: The radial integral does not converge, indicating the failure of four-Fermi theory as we discussed previously.
\item $\ell\geq 2$: In this case, the integral in Eq.~\eqref{eq:2l-3} does converge and four-Fermi theory is suitable for such states. The result is
\begin{equation}
\frac{\alpha^2}{m_e m _Z ^4 a_0^6}\int_0^{\infty} \D \eta \  \eta^{2\ell-3}\exp(-n_{sup}\eta)\sim m_e \alpha ^8 \left(\frac{m_e}{m_Z}\right)^4 ,
\end{equation}
where we have ignored some $\mathcal{O}(1)$ constants that depend on $\ell$. 
\end{enumerate}

In Table~\ref{Tab:comparison},we compare the tree-level and loop-level matrix elements for different values of $\ell$.
For $\ell=2$, the tree-level matrix element behaves as $\alpha^9 \left(m_e/m_Z\right)^6$, while the loop matrix element goes as $\alpha^8 \left(m_e/m_Z\right)^4$. Thus, naively, for $\ell=2$,
\beq
\frac{\mathcal{M}_{\text{tree}}}{\mathcal{M}_{\text{loop}}} \sim \alpha \left(\frac{m_e}{m_Z} \right)^2 \approx 10^{-13}.
\eeq
In other words, the effect of the tree-level potential is much smaller than the effect of the loop-level potential for $\ell\geq 2$. If we only care about powers of $\alpha$ and $m_e/m_Z$, then our calculations suggest that the effect of the loop remains the same as $\ell\geq 2$, i.e,  $\sim  \alpha ^8 \left(m_e/m_Z\right)^4$, but the powers in $\alpha$ and $m_e/m_Z$ in the tree-level effect increase with $\ell$, rendering it much smaller. Thus, to isolate the effects of the loop, we need to consider states for which $\ell\geq 2$.
\begin{table}
\begin{center}
\begin{tabular}{|c|c|c|}
\hline 
$\ell$ & From $ V_{PNC}^{\text{tree}}$ &  $ V_{PNC}^{\text{loop}}$ \\ 
\hline 
$\ell=0$ & $\sim  \alpha ^5 \left(\frac{m_e}{m_Z}\right)^2$ & does not converge \\ 
\hline 
$\ell=1$ & $\sim  \alpha ^7 \left(\frac{m_e}{m_Z}\right)^4$ & does not converge \\ 
\hline 
$\ell\geq 2$ & $\sim  \alpha ^{2l+5} \left(\frac{m_e}{m_Z}\right)^{2\ell+2}$ & $\sim  \alpha ^8 \left(\frac{m_e}{m_Z}\right)^4$ \\
\hline 
\end{tabular} 
\end{center}
\caption{Tree-level and loop-level matrix elements for different values of $\ell$}
\label{Tab:comparison}
\end{table}

\section{A sample calculation}\label{sec:sample_calc}
Note that while calculating matrix elements of the potential between two states of definite orbital angular momenta, we took the lesser of the two to be $\ell$ and the higher to be $\ell'$. In order for the matrix element to converge in the four-Fermi approximation, we need $\ell \geq 2$. In other words, the lowest angular momentum state that we can work with in a matrix element calculation is $\ell =2$.
Based on this, we explore parity-violating corrections to some of the $\ell=3$ states of the hydrogen atom. Because of a parity non-conserving potential, $\ell=3$ states can only mix with $\ell=2$ and $\ell=4$ states, which both satisfy the convergence criterion. At the same time, the wave function of these states falls to zero at the origin faster than the $s$ or the $p$ states, and so one could hope that, in states with $\ell=3$, some parity-violation effect can be brought about predominantly by the neutrino loop instead of by the $Z$-interaction. We emphasize here that we could not have chosen $\ell =2$ states for this task, because these states mix with $\ell=1$ states when there is parity violation, which do not satisfy the convergence criterion that $\ell \geq 2$.

As discussed in Sec.~\ref{intro}, parity violation in atoms is measured in optical rotation experiments, wherein the degree of rotation of the plane of polarization of light is proportional to $R$ defined in Eq.~\eqref{eq:angle_rot}. In this section, we study a particular interference process between two eigenstates of hydrogen and its effect on the plane of polarization of linearly-polarized incident light on a hydrogen sample.

Note that $M1$ transitions between states of different principal quantum number $n$ do not occur in hydrogen because of the orthogonality of states with different $n$. To observe this effect, we therefore need to look for two states with the same parity and the same principal quantum number. To this end, we consider the following states of definite $n,f,m_f,j,\ell$ in the notation $\ket{n,f,m_f,j,\ell}
$:
\begin{eqnarray}
\ket{A}=\ket{4,3,3,5/2,3} &\equiv& 4F_{5/2,F=3},\\
\ket{B}=\ket{4,3,3,7/2,3} &\equiv& 4F_{7/2,F=3} ,\\
\ket{\Delta}=\ket{4,3,3,5/2,2} &\equiv& 4D_{5/2,F=3}
\end{eqnarray}
$\ket{A}$ and $\ket{B}$ are eigenstates of $\hat{F}^2$ which, in the presence of the neutrino potential, mix with all other states with $f=3$ and $m_f =3$ to form a true energy eigenstate of hydrogen. Before adding the neutrino potential, these states have the same $\ell$ and hence there can be an $M1$ transition between them, but no $E1$ transition. However, once these states are corrected by the neutrino potential, the resulting eigenstates can have both $E1$ and $M1$ transitions between them because of the small parity violating correction, from which we can calculate $R$, as in Eq.~\eqref{eq:angle_rot}.

Consider now the state $\ket{\Delta}$. This state has different parity than the two base states $\ket{A}$ and $\ket{B}$ while having the same $f$ and $m_f$ quantum numbers and, hence, can mix with them. Before we proceed, we note that other states with the same values of $f$ and $m_f$, such as $\ket{5,3,3,7/2,4}$ for instance, mix very weakly with our base states because the quantum number $n$ puts these states much farther away in energy than $\ket{\Delta}$. We therefore ignore the contribution of these states in the perturbation expansion. Lastly, we must keep in mind that the matrix element of a parity-violating operator between states with the same parity is zero. Therefore, the base states do not get any corrections from each other since they have the same $\ell=3$.

Our aim is to compute \beq \frac{\mel{\text{Electric Dipole}}{A'}{B'}}{\mel{\text{Magnetic Dipole}}{A'}{B'}}\approx\frac{\mel{\text{Electric Dipole}}{A'}{B'}}{\mel{\text{Magnetic Dipole}}{A}{B}} \label{elmag} \eeq
where $\ket{A'}$ and $\ket{B'}$ are the true eigenstates of hydrogen, obtained from $\ket{A}$ and $\ket{B}$ using the perturbation expansion as in Eq.~\eqref{pert}. For details of the calculation, see appendix \ref{app:calc}.
The approximation in Eq.~\eqref{elmag} holds because the selection rules permit magnetic transitions to occur between states of the same parity, so perturbative corrections, which are much smaller than the unperturbed transition amplitude, can be ignored.

Using the electric and magnetic dipole moment operators (details in the appendix), we compute the inner products by performing the integrals involving the hydrogen atom wavefunctions. We define a small parameter $\nu_i$ by:
\beq
\nu_i \equiv \frac{1}{\alpha} \frac{m_{\nu_i}}{m_e}
\eeq
The final result, up to leading order in  $\nu_i$ is
\begin{eqnarray}
R  &=&\frac{-7  \alpha  m_e^3 m_p G_A G_F^2\left(-\frac{1}{4}+   s_W^2 +\frac{1}{2}|U_{ei}|^2\right)}{302778777600 \pi ^3 g_p (29 g_p m_e-21609000 m_p)} \\
 & & \times \left[(24335 g_p m_e-17503290000 m_p)+\nu_i^2(3858 g_p m_e+84015792000 m_p)\right] + \mathcal{O}(\nu_i^4), \nonumber
\end{eqnarray}
where there is an implicit sum over the neutrino flavor $i$.
Using the standard values of the quantities above, we find
\begin{eqnarray}
R = \Im\left(\frac{E1_{PV}}{M1}\right) \approx \left(-\frac{1}{4}+  s_W^2 +\frac{1}{2}|U_{ei}|^2\right)\left(-7.7\times10^{-33}+3.7\times10^{-32} \nu_i^2\right). \label{eq:finres}
\end{eqnarray}
The result above shows that the leading-order contribution to $R$ is a number of order $\mathcal{O}(10^{-32})$. The next-to-leading-order term depends on the neutrino mass through the parameter $\nu_i $. Using current experimental bounds on the neutrino mass $(m_{\nu} < 0.12 \  \text{eV})$, we see that the next-to-leading-order term has a magnitude of $\mathcal{O}(10^{-41})$ radians. 

Upon completing the calculation of the specific rotation here, let us provide some perspective on the result. We first compare the value of $R$ obtained from a neutrino loop diagram to the typical values obtained from a $Z$ diagram. To this end, we choose the states $\ket{2,1,1,\frac{1}{2},1}$ and $\ket{2,1,1,\frac{3}{2},1}$. Both of these states have $f=1$, and $\ell=1$ and both  are corrected by the state $\ket{2,1,1,\frac{1}{2},0}$.  Note that we have picked low $\ell$ states since we show in Sec.~\ref{sec:effects} that the $Z$ diagram dominates for such states. The precise choice of states is not completely without motivation: We have picked $p$-wave states with $n=2$ because these states experience relatively large corrections from the $s$-wave states with the same principal quantum number. Had we picked $s$-wave states with $n=1$, the corrections  would be rather small. This is because they would come from $\ell=1$ states which are much farther separated in energy, since the $n=1$ shell does not possess any $\ell=1$ states. 

We repeat the process outlined in this section with only the first term in Eq.~\eqref{vtree} for these two states, and obtained
\begin{eqnarray}
R=\Im\left(\frac{E1_{PV}}{M1}\right) = \frac{27 g^2 m_p [g_p m_e (4323 \eta_Z +1730)-162 m_p(2 \eta_Z+1)]}{6904 \pi \cos^2\theta_W \alpha ^3 g_p m_e (\eta_Z+1)^3 (865 g_p m_e-81 m_p)},
\end{eqnarray}
where $\eta_Z = m_Z (m_e \alpha)^{-1} \gg 1$. 
After plugging in the standard numerical values, we have
\begin{eqnarray}
R=\Im\left(\frac{E1_{PV}}{M1}\right) \sim 10^{-10}.
\end{eqnarray}
It turns out, therefore, that the $Z$-diagram gives an optical rotation for $\ell =1$ states that is about $10^{22}$ times larger than the optical rotation obtained from the neutrino loops for the higher $\ell = 3$ states.

\section{Final remarks}\label{sec:conclusion}
From the results in Sec.~\ref{sec:sample_calc}, it is clear that the measurement of optical rotation due to the neutrino loop is extremely challenging given the resolutions we can achieve today. In that regard, there is another obstacle in the path to measuring this effect -- that of statistical suppression. Since we are looking at high-$\ell$ states, they necessarily occur at high $n$, which means that these are high-energy states and are thermally suppressed. We saw earlier that, for the lower energy states, the parity-violating interaction via the Z exchange dominates over the neutrino process. Hence, at low temperatures, the chances of isolating the neutrino-mediated transition are pretty low. 

Nonetheless, this calculation, performed for other systems, could lead to somewhat larger quantities and the next step would most likely be an application of this idea to many-electron atoms, beyond the simple hydrogen case. Multi-electron atoms are important to explore particularly because the matrix elements in these atoms are amplified by an additional $Z^3$ factor \cite{Bouchiat:1974zz}, $Z$ being the atomic number of the heavy atom in question. The $Z^3$ amplification is only present when one considers low-$\ell$ states of heavy atoms - one factor of $Z$ comes in through the weak nuclear charge and the other two factors appear out of the relativistic behavior of low-$\ell$ electrons near the nucleus. It might be worthwhile to try to explore the long-range parity violation in heavier atoms, but it is still very unlikely that we may be able to isolate the effect of the neutrino loop since the $Z^3$ amplification factor acts on both the tree level and loop level effects.

To conclude, we highlight the merits and demerits of the calculation: Although the effects of the neutrino force on the hydrogen atom are extremely small to measure in an experiment, the neutrino force is the largest long-range parity-violating force there is. 
 
\section{Acknowledgments}
We thank Maxim Pospelov for pointing out a mistake in the first version of
the paper that resulted in a revised section~\ref{sec:photon-pen}.
The work of YG is supported in part by the NSF grant PHY1316222. The research of WT is supported by the College of Arts and Sciences of Loyola University Chicago.
 
\begin{appendix}
\section{Calculation of the parity violating force between the electron and the proton}\label{app:pnc}
Our approach here closely follows the methodology of \cite{Feinberg:1989ps}.
For the sake of simplicity, we start by assuming just one flavor for the neutrino. In that case we find the following four-Fermi operator for two fermions of type $\psi$ and two neutrinos  by summing over the $Z$ and $W$ diagrams:
\begin{equation}
\mathcal{O}_4 = -\frac{G_F}{\sqrt{2}}[\bar{\psi}\gmu (a^{\psi} - b^{\psi} \gfive)\psi][\bar{\nu}\gMu(1-\gfive)\nu],
\end{equation}
where $a^{\psi}$ and $b^{\psi}$ are the effective couplings to the $Z$ as defined in Eqs.~\eqref{eq:abijl} and \eqref{eq:abijn}. They depend on the particular fermion in question, depending on whether the $W$ exchange contributes, the $Z$ exchange contributes, or both. 

The two-neutrino potential can be calculated by a double insertion of this operator, and the evaluation of the resulting amplitude, and by taking the Fourier transform of the amplitude. The Feynman diagram that is relevant is given in Fig.~\ref{fig:2nu}. The corresponding matrix element is given by
\begin{equation}
i \mathcal{M} = -\frac{(-iG_F) ^2}{2}\bar{e}\bar{N}\left[\Gamma ^e_\mu \Gamma ^N _{\nu}\right] \int \frac{\D ^4 k \D ^4 k' }{(2\pi)^4}\delta^4(q-k-k')\text{Tr}\left[i\Gamma ^{\mu} \frac{i(-\slashed{k'}+m)}{k'^2-m^2}i\Gamma ^{\nu}\frac{i(\slashed{k}+m)}{k^2-m^2}\right]eN .
\end{equation}
Here, $\Gamma_{\mu} ^ f = \gMu (a_f -b_f \gfive)$, with $a_f$ and $b_f$ depending on the type of fermion in question. $N$ stands for nucleus, which in our case is just the proton. We can write the matrix element as $i\mathcal{M} = \bar{e}\bar{N} iF eN$, where:
\begin{equation}
F = -i\frac{G_F ^2}{2}\left[\Gamma ^e_\mu \Gamma ^N _{\nu}\right] \int \frac{\D ^4 k \D ^4 k' }{(2\pi)^4}\delta^4(q-k-k')\text{Tr}\left[\Gamma ^{\mu} \frac{(-\slashed{k'}+m)}{k'^2-m^2}\Gamma ^{\nu}\frac{(\slashed{k}+m)}{k^2-m^2}\right] .
\end{equation}
We then evaluate the trace, and consider only the symmetric part, since the antisymmetric part is odd in $k$, and hence evaluates to $0$ in the loop integral,
\begin{eqnarray}
F &=& i\frac{G_F ^2}{2}\left[\Gamma ^e_\mu \Gamma ^N _{\nu}\right] 2\text{Tr}\left[\gmu\grho\gnu\gsig\right] \int \frac{\D ^4 k \D ^4 k' }{(2\pi)^4}\delta^4(q-k-k')\frac{k_{\sigma}k'_{\rho}}{(k^2-m^2)(k'^2-m^2)}, \\ 
 &=&  i\frac{G_F ^2}{2}\left[\Gamma ^e_\mu \Gamma ^N _{\nu}\right]C^{\mu \nu; \rho \sigma} I_{\sigma \rho} \nonumber,
\end{eqnarray}
where, 
\begin{eqnarray}
C^{\mu \nu; \rho \sigma} &\equiv& 2\text{Tr}\left[\gmu\grho\gnu\gsig\right], \\
I_{\sigma \rho} &\equiv& \int \frac{\D ^4 k \D ^4 k' }{(2\pi)^4}\delta^4(q-k-k')\frac{k_{\sigma}k'_{\rho}}{(k^2-m^2)(k'^2-m^2)} = A'g_{\rho\sigma} + B'q_{\sigma}q_{\rho}.
\end{eqnarray}
We can therefore write, after contracting $I_{\sigma \rho}$ with $g_{\rho\sigma}$ and $q_{\sigma}q_{\rho}$ respectively,
\begin{eqnarray}
4A' + B't &=& \int \frac{\D ^4 k \D ^4 k' }{(2\pi)^4}\delta^4(q-k-k')\frac{k.k'}{(k^2-m^2)(k'^2-m^2)} \equiv J_0,\\
A't + B't^2 &=& \int \frac{\D ^4 k \D ^4 k' }{(2\pi)^4}\delta^4(q-k-k')\frac{(q.k)(q.k')}{(k^2-m^2)(k'^2-m^2)} \equiv J_1,
\end{eqnarray}
where $t$ is the Mandelstam variable.\\

To calculate the force, we find the discontinuity in the matrix element across the branch cut in the complex $t$ plane, using the Cutkosky cutting rules, which yields
\begin{eqnarray}
\!\!\!\!\!\!\!\!\!\!\! \tilde{J_0} &=& -\frac{1}{(2\pi)^2}\int \D ^4 k \D ^4 k' \delta^4(q-k-k')\theta (k^0)\theta(k'^0)\delta(k^2-m^2)\delta(k'^2 -m^2)(k\cdot k'), \\
\!\!\!\!\!\!\!\!\!\!\! \tilde{J_1} &=& -\frac{1}{(2\pi)^2}\int \D ^4 k \D ^4 k' \delta^4(q-k-k')\theta (k^0)\theta(k'^0)\delta(k^2-m^2)\delta(k'^2 -m^2)(k\cdot q)(k'\cdot q).
\end{eqnarray}
Here, the tilde denotes the discontinuity of a quantity across the branch cut. Writing
$$C^{\mu \nu; \rho \sigma}\left(A'g_{\rho\sigma} + B'q_{\sigma}q_{\rho}\right)= A g^{\mu\nu} + B'q^{\mu}q^{\nu}, $$
we obtain
\begin{equation}
A\, = \,-8\,(2A'+B't),\quad B\, = \, 16\,B' .
\end{equation}
We have then,
\begin{eqnarray}
F &=& i\frac{G_F ^2}{2}\left(\Gamma ^e \cdot \Gamma ^N A + q^{\mu}q^{\nu} \Gamma ^e _{\mu} \Gamma ^N _{\nu} B\right),\\
\tilde{F}& =& i\frac{G_F ^2}{2}\left(\Gamma ^e \cdot \Gamma ^N \tilde{A} + q^{\mu}q^{\nu} \Gamma ^e _{\mu} \Gamma ^N _{\nu} \tilde{B}\right). \label{maineq}
\end{eqnarray}
What we need is to calculate the discontinuity in the matrix element since the spectral function $\rho$ is given by:
\begin{equation}
\rho = \frac{\tilde{\mathcal{M}}}{2i} .
\end{equation}
We evaluate the integrals above in the CM frame of momentum transfer, i.e, the frame where $q = (\sqrt{t},0,0,0)$,
and hence  
$k = (\omega, \vec{k}),$ 
$k' = (\omega', -\vec{k}).$

Performing the integrals, in the case of equal masses of the neutrino in both propagators of the loop, we have
\begin{eqnarray}
\tilde{J_0} &=& -\frac{1}{16\pi}\sqrt{1-\frac{4m^2}{t}}(t-2m^2), \quad
\tilde{J_1} \,=\, -\frac{t^2}{32\pi}\sqrt{1-\frac{4m^2}{t}} .
\end{eqnarray}
Which yields
\begin{eqnarray}
\tilde{A'}&=&-\sqrt{1-\frac{4m^2}{t}}\left(\frac{t-4m^2}{96\pi}\right), \quad
\tilde{B'} \,=\, -\sqrt{1-\frac{4m^2}{t}}\left(\frac{t^2+2m^2 t}{32\pi t^2}\right),
\end{eqnarray}
and translates to
\begin{eqnarray}
\tilde{A}&=&\sqrt{1-\frac{4m^2}{t}}\left(\frac{t-m^2}{3\pi}\right), \quad \tilde{B} \,=\, -\sqrt{1-\frac{4m^2}{t}}\left(\frac{1+\frac{2m^2}{t}}{3\pi}\right).
\end{eqnarray}

We now need to deal with Eq.~\eqref{maineq}, and evaluate the spinor products in the non-relativistic limit. For the purpose of calculating the velocity-dependent terms in the potential, it is necessary to evaluate the spinors upto first order in momentum $\vec{p}$. This calculation seems most convenient in the Pauli-Dirac basis where the non-relativistic limit is much easier to work with. In the Pauli-Dirac basis, a Dirac spinor is given by
\begin{equation}
\uspinor{s}{p} = \sqrt{p^0+m}\begin{pmatrix}
\xi_s \\ \frac{\vec{\sigma}\cdot\vec{p}}{p^0+m}\xi_s
\end{pmatrix} .
\end{equation} 
The gamma matrices, in this basis, are given by
\begin{eqnarray}
\gamma^0 &=& \begin{pmatrix}
\textbf{1}& 0\\0 & -\textbf{1}
\end{pmatrix}, \quad
\gamma^i = \begin{pmatrix}
0 &\sigma^i\\-\sigma^i & 0
\end{pmatrix}, \quad
\gamma^5 = \begin{pmatrix}
0 & \textbf{1}\\ \textbf{1} & 0
\end{pmatrix} .
\end{eqnarray}
In the non-relativistic limit, $p^0+m \to 2m$, and therefore, for the electron,
\begin{equation}
\uspinor{s}{p} \approx \sqrt{2m}\begin{pmatrix}
\xi_s \\ \frac{\vec{\sigma}\cdot\vec{p}}{2m}\xi_s 
\end{pmatrix} = \sqrt{2m}\begin{pmatrix}
\xi_s \\ \frac{\vec{\sigma}\cdot\vec{v}}{2}\xi_s  
\end{pmatrix} ,
\end{equation}
where $\xi_s$ is a 2-component vector that encodes the spin state.
For the nucleus, which has mass $M \gg m$, we can write
\begin{equation}
\uspinor{r}{p} \approx \sqrt{2M}\begin{pmatrix}
\xi_r \\ 0 
\end{pmatrix} .
\end{equation}
We  use the above approximation for evaluating $\mathcal{M}$. Our plan is to evaluate the integral that gives us the long-range potential from the spectral function.

The $q_{\mu}q_{\nu}$ term does not give a parity violating term when evaluated explicitly using spinors. Thus, we only need to evaluate the $\Gamma ^e \cdot \Gamma ^N$ term. We suppress writing the spin states $\xi$, and assume that the incoming and outgoing electrons have 3-momenta $\vec{p}$ and $\vec{p} \ '$ respectively, while the incoming and outgoing nuclei have 3-momenta $\vec{k}$ and $\vec{k}'$ (note, as usual that $q = p-p'=k'-k$, let us not confuse the $k$'s here with the integration variables used before --- those $k$'s have no relevance in the upcoming discussion).
To compute the leading radial dependence of the potential, we need the spin and momentum independent parity conserving term in $F$. This is found to be $2i m_e M a_e a_N G_F ^2A$. The discontinuity in the matrix element for the spin-independent part is
\begin{align}
\tilde{\mathcal{M}} &= 2m_eMia_e a_N G_F ^2\tilde{A} = 2m_eMia_e a_N G_F ^2 \sqrt{1-\frac{4m^2}{t}}\left(\frac{t-m^2}{3\pi}\right) .
\end{align}
The spectral function is therefore (ignoring the spin states)
\begin{eqnarray}
\rho(t) &=& \frac{\tilde{\mathcal{M}}}{2i}= m_eMa_e a_N G_F ^2 \sqrt{1-\frac{4m^2}{t}}\left(\frac{t-m^2}{3\pi}\right).
\end{eqnarray}
Thus, the spin-independent parity conserving potential is given by the formula
\begin{eqnarray}
V(r)&=& \frac{1}{16\pi^2 m_eM r}\int_{t_0} ^{\infty} \D t \ \rho (t) e^{-\sqrt{t}r}, \\
&=& \frac{m_eMa_e a_N G_F ^2}{16\pi^2 m_eM r}\int_{4m^2}^{\infty} \D t \  e^{-\sqrt{t}r}  \sqrt{1-\frac{4m^2}{t}}\left(\frac{t-m^2}{3\pi}\right), \nonumber \\
&=&\frac{a_e a_N G_F ^2}{4\pi^3 }\frac{m^3 K_3 (2mr)}{r^2},\nonumber \\
&=& a_e a_N V_{\nu \nu}(r),\nonumber
\end{eqnarray}
where $V_{\nu \nu}(r)$ is given in Eq.~\eqref{Massive_nu} (the Dirac case).

We also calculate the parity violating parts, as below:
\begin{align}
\frac{\bar{e}\bar{N}\Gamma ^e . \Gamma ^N  eN}{4m_eM} &\supset a_N b_e\left(\frac{1}{2m_e}+\frac{1}{2M}\right)\vec{\sigma}_e\cdot\vec{q}-a_e b_N\left(\frac{1}{2m_e}+\frac{1}{2M}\right)\vec{\sigma}_N\cdot\vec{q}+\frac{a_eb_N}{m_e}\vec{\sigma}_N\cdot\vec{p} \nonumber \\
& -\frac{a_Nb_e}{m_e}\vec{\sigma}_e\cdot\vec{p} +i\left(\frac{a_e b_N}{2m_e}+\frac{a_N b_e}{2M}\right)\vec{\sigma}_e\cdot (\vec{\sigma}_N\times \vec{q}). \label{eq:FPV0}
\end{align}
The  parity violating parts of $F$ are therefore given by:
\begin{align}
\frac{F}{2 i G_F^2 m_eM} &\supset \left[a_N b_e\left(\frac{1}{2m_e}+\frac{1}{2M}\right)\vec{\sigma}_e\cdot\vec{q}\right]A-\left[a_e b_N\left(\frac{1}{2m_e}+\frac{1}{2M}\right)\vec{\sigma}_N\cdot\vec{q}\right]A \nonumber\\
&+\left[\frac{a_eb_N}{m_e}\vec{\sigma}_N\cdot\vec{p}\right]A -\left[\frac{a_Nb_e}{m_e}\vec{\sigma}_e\cdot\vec{p}\right]A \nonumber \\
& +i\left[\left(\frac{a_e b_N}{2m_e}+\frac{a_N b_e}{2M}\right)\vec{\sigma}_e\cdot (\vec{\sigma}_N\times \vec{q})\right]A .\label{eq:FPV}
\end{align}

$V_{\nu \nu}(r)$ is basically the Fourier transform of the spin-independent part of the matrix element $\mathcal{M}$, i.e, it can be thought of as the Fourier transform of $A$, upto the non-relativistic normalization of the Dirac spinors. But observe that the spin-dependent part of the matrix element is obtained by multiplying the spin independent term $A$ to the terms in Eq. \eqref{eq:FPV0}. Thus, to obtain the spin dependent parts of the potential, we need to take the Fourier transforms of quantities such as $(\vec{\sigma}\cdot \vec{q}) A$ and so on.  In essence, we replace $\vec{q}$ 's by gradients. 


Let us look at the particular case of the hydrogen atom. We incorporate flavor mixing as in sec. \ref{sec:pncforce}, and get the couplings $a^e_{ii}$, $b^e_{ii}$, $a^p_{ii}$ and $b^p_{ii}$ as in Eq.~\eqref{Eq:couplingsmix}. 

For sake of cleanliness, below we drop one index $i$ from the above couplings, since no sum is assumed anyway.
The analog of Eq.~\eqref{eq:FPV0} in the hydrogen atom is therefore (the Hermitian conjugate is implicitly added)
\begin{align*}
\frac{\bar{e}\bar{P}\Gamma ^e . \Gamma ^P  eP}{4m_em_p} &= a^p_i b^e_i\left(\frac{1}{2m_e}+\frac{1}{2m_p}\right)\vec{\sigma}_e\cdot\vec{q}-a^e_i b^p_i\left(\frac{1}{2m_e}+\frac{1}{2m_p}\right)\vec{\sigma}_p\cdot\vec{q}+\frac{a^e_ib^p_i}{m_e}\vec{\sigma}_p\cdot\vec{p}  \nonumber \\
& -\frac{a^p_ib^e_i}{m_e}\vec{\sigma}_e\cdot\vec{p}+i\left(\frac{a^e_i b^p_i}{2m_e}+\frac{a^p_i b^e_i}{2m_p}\right)\vec{\sigma}_e\cdot(\vec{\sigma}_p\times \vec{q}),\\
&\approx \frac{a^e_i b^p_i}{2m_e}\left[2\vec{\sigma}_p\cdot\vec{p} -\vec{\sigma}_p\cdot\vec{q}+i\vec{\sigma}_e\cdot(\vec{\sigma}_p\times \vec{q})\right],\\
&= \frac{G_A}{2m_e}\left(-\frac{1}{4}+  \sin ^2 \theta_W +\frac{1}{2}|U_{ei}|^2\right)\left[2\vec{\sigma}_p\cdot\vec{p} -\vec{\sigma}_p\cdot\vec{q}+i\vec{\sigma}_e\cdot(\vec{\sigma}_p\times \vec{q})\right].
\end{align*}
Here, we used the fact that $\sin ^2 \theta_W \approx 0.23$ so that $a^p_i \sim 0$ and that $m_e \ll m_p$.
The parity-violating potential that comes out of this with a Fourier transform is given by (we remember to add in the Hermitian conjugate and implicitly sum over $i$)

\begin{eqnarray}
V_{PNC}^{\text{loop}} &=& \frac{G_A}{m_e}\left(-\frac{1}{4}+  \sin ^2 \theta_W +\frac{1}{2}|U_{ei}|^2\right)\left[(2\vec{\sigma}_p\cdot\vec{p})V_{\nu_i \nu_i}(r) +\vec{\sigma}_e\cdot(\vec{\sigma}_p\times \vec{\nabla})V_{\nu_i \nu_i}(r)\right], \nonumber \\
&=&\frac{G_A}{m_e}\left(-\frac{1}{4}+  \sin ^2 \theta_W +\frac{1}{2}|U_{ei}|^2\right)\left[(2\vec{\sigma}_p\cdot\vec{p})V_{\nu_i \nu_i}(r) +(\vec{\sigma}_e\times\vec{\sigma}_p)\cdot \vec{\nabla}V_{\nu_i \nu_i}(r)\right] \label{vpnc}
\end{eqnarray}

\section{Details of the calculation in Sec. \ref{sec:sample_calc}}\label{app:calc}
In Sec.~\ref{sec:sample_calc}, we computed $R$, for the $E1$ and $M1$ transitions between the ``base'' states $\ket{A}$ and $\ket{B}$. Both of these states were corrected by the ``correction state'' $\ket{\Delta}$. Other corrections were ignored because they are much smaller than the correction due to $\ket{\Delta}$.

Using the machinery of angular-momentum addition, we can write
\begin{eqnarray}
\ket{A}=\ket{4,3,3,5/2,3} &\equiv& -\frac{1}{\sqrt{7}}\psi_{432}\ket{\uparrow \uparrow}+\sqrt{\frac{6}{7}}\psi_{433} \ket{\downarrow \uparrow},\\
\ket{B}=\ket{4,3,3,7/2,3} &\equiv& -\frac{1}{2}\sqrt{\frac{3}{7}}\psi_{432} \ket{\uparrow \uparrow} + \frac{1}{2}\sqrt{\frac{7}{2}}\psi_{433}\ket{\uparrow \downarrow}-\frac{1}{2\sqrt{14}}\psi_{433}\ket{\downarrow \uparrow}, \nonumber \\
\ket{\Delta}=\ket{4,3,3,5/2,2} &\equiv& \psi_{422}\ket{\uparrow \uparrow}, \nonumber
\end{eqnarray}
where $\psi_{nlm}$ are the unperturbed energy eigenstates of hydrogen, given by
\begin{equation}
\psi_{nlm} = \bk{r,\theta,\phi}{nlm}=\sqrt{\left(\frac{2}{na_0} \right) ^3 \frac{(n-l-1)!}{2n[(n+l)!]^3}}e^{-r/na_0} \left[L^{2l+1}_{n-l-1}(2r/na_0) \right]Y^m _l(\theta, \phi).
\end{equation}

Using these three states, we can write the corrected states in the spirit of Eq.~\eqref{pert} as:
\begin{eqnarray}
\ket{A'} &=& \ket{A} + \frac{\mel{V_{PNC}}{\Delta}{A}}{E_A-E_{\Delta}} \ket{\Delta} +\cdots = \ket{A} + C_{A\Delta} \ket{\Delta} +\cdots, 
\end{eqnarray}
where $C_{A\Delta}$ is the correction coefficient. Similarly,
\begin{eqnarray}
\ket{B'} &=& \ket{B} + \frac{\mel{V_{PNC}}{\Delta}{B}}{E_B-E_{\Delta}} \ket{\Delta} +\dots = \ket{B} + C_{B\Delta} \ket{\Delta} +\dots
\end{eqnarray}
In the end, we add the contributions from both terms in the potential. Our states therefore become
\begin{eqnarray}
\ket{A'}=\ket{B} + (C^{sc}_{A\Delta} +C^{h}_{A\Delta} )\ket{\Delta} +\cdots,\\
\ket{B'}=\ket{B} + (C^{sc}_{B\Delta}+C^{h}_{B\Delta} ) \ket{\Delta} +\cdots.
\end{eqnarray}
Here $C^{sc}$ is the correction coefficient for the spin-cross term alone, while $C^{h}$ is the coefficient for the ``helicity'' term alone.

Using the two terms in $V_{PNC}^{\text{loop}} (r)$, we compute the corrections up to second order in the small parameter $\nu_i$. To calculate the energy differences between the states, we use Eq. \eqref{Henergy}. We obtain ($s_W \equiv \sin\theta_W$) 
\begin{eqnarray}
\!\!\!\!\!\!\!\!\!\! C^{\text{sc}}_{A\Delta} &=& i\frac{G_A G_F^2m_p m_e ^3 \alpha^2}{\pi^3 g_p}\left(-\frac{1}{4}+  s_W^2 +\frac{1}{2}|U_{ei}|^2\right)\left(\frac{21\sqrt{7}\nu_ i^2}{10816}- \frac{35\sqrt{7}}{64896}\right),\\
\!\!\!\!\!\!\!\!\!\! C^{\text{sc}}_{B\Delta} &=& i\frac{G_A G_F^2m_p m_e ^4 \alpha^2}{\pi^3 (29g_p m_e-21609000m_p)}\left(-\frac{1}{4}+   s_W^2 +\frac{1}{2}|U_{ei}|^2\right) \times\left(-\frac{7\sqrt{\frac{7}{3}}\nu_i^2}{64}+ \frac{35\sqrt{\frac{7}{3}}}{1152}\right),\\
\!\!\!\!\!\!\!\!\!\! C^{\text{h}}_{A\Delta} &=& \left(-\frac{1}{4}+   s_W^2 +\frac{1}{2}|U_{ei}|^2\right)\frac{7 i \sqrt{7} \alpha ^2 m_e^3 \left(36 \nu_i ^2-5\right) m_p G_A G_F^2}{129792 \pi ^3 g_p},\\
\!\!\!\!\!\!\!\!\!\! C^{\text{h}}_{B\Delta} &=& \left(-\frac{1}{4}+  s_W^2 +\frac{1}{2}|U_{ei}|^2\right)\frac{7 i \sqrt{\frac{7}{3}} \alpha ^2 m_e^4 \left(1122 \nu_i ^2-115\right) M G_A G_F^2}{27648 \pi ^3 (29 g_p m_e-21609000 m_p)}.
\end{eqnarray}

We are interested in the ratio between the electric and magnetic dipole moment matrix elements for the states $\ket{A'}$ and $\ket{B'}$. These two transition matrix elements have the same dependence on the magnetic quantum numbers in hydrogen, and so the ratio is independent of the orientation of the atom. As such, in our calculations, we only look at the magnetic and electric dipole moments along the $z$ direction,
$$ \hat{P} = -ez = -(4\pi\alpha)^{1/2} r \cos\theta,$$
$$\hat{M} = \frac{e}{2m_e}(\hat{L}_z+2 \hat{S}_z)  = \frac{(4 \pi \alpha)^{1/2}}{2m_e}(\hat{L}_z+2 \hat{S}_z).$$
Using this form of electric and magnetic dipole moment operators in Eq.~\eqref{elmag} leads to the final result in Eq. \eqref{eq:finres}.

\end{appendix}

\bibliography{APV}
\bibliographystyle{apsrev4-1}

\end{document}